\documentclass{webofc}
\usepackage[varg]{txfonts}   % Web of Conferences font
\newcommand{\beq}{\begin{eqnarray}}
\newcommand{\eeq}{\end{eqnarray}}
\newcommand{\Sp}[1]{S\!p(#1)}
\newcommand{\Spfour}{\Sp{4}}
\newcommand{\SptwoN}{\Sp{2N}}
\newcommand{\SU}[1]{S\!U(#1)}
\newcommand{\SUfour}{\SU{4}}
\newcommand{\SUN}{\SU{N}}
\renewcommand{\Re}{\mathrm{Re}}
\begin{document}
\title{Sp(4) gauge theories and beyond the standard model physics}
%
% subtitle is optional

\author{\firstname{Biagio}
  \lastname{Lucini}\inst{1,2}\fnsep\thanks{Speaker;
    \email{b.lucini@swansea.ac.uk}} \and
        \firstname{Ed} \lastname{Bennett}\inst{2} \and
        \firstname{Jack} \lastname{Holligan}\inst{3,4} \and
        \firstname{Deog Ki} \lastname{Hong}\inst{5} \and
        \firstname{Ho} \lastname{Hsiao}\inst{6} \and
        \firstname{Jong-Wan} \lastname{Lee}\inst{5,7} \and
        \firstname{C.-J. David} \lastname{Lin}\inst{6,8,9} \and
        \firstname{Michele} \lastname{Mesiti}\inst{2} \and
        \firstname{Maurizio} \lastname{Piai}\inst{3} \and
        \firstname{Davide} \lastname{Vadacchino}\inst{10}
}

\institute{
Department of Mathematics, Swansea University, Fabian Way, SA1 8EN
Swansea, Wales, UK  
\and
Swansea Academy of Advanced Computing, Swansea University, Fabian Way, SA1 8EN Swansea, Wales, UK
\and
Department of Physics, Swansea University, Singleton Park, SA2 8PP Swansea, Wales, UK
\and
Department of Physics, University of Maryland, College Park, Maryland, USA, 20742
\and
Department of Physics, Pusan National University, Busan 46241, Korea
\and
Institute of Physics, National Yang Ming Chiao Tung University, 1001 Ta-Hsueh Road, Hsinchu 30010, Taiwan 
\and
Extreme Physics Institute, Pusan National University, Busan 46241, Korea
\and
Center for High Energy Physics, Chung-Yuan Christian University,
Chung-Li 32023, Taiwan 
\and
Centre for Theoretical and Computational Physics, National Yang Ming Chiao Tung University, 1001 Ta-Hsueh Road, Hsinchu 30010, Taiwan
\and
School of Mathematics and Hamilton Mathematics Institute, Trinity
College, D02 PN40 Dublin 2, Ireland
          }

\abstract{%
We review numerical results for models with gauge group $\SptwoN$,
discussing the glueball spectrum in the large-$N$ limit, the quenched meson spectrum
of $\Spfour$ with Dirac fermions in the fundamental and in the antisymmetric representation
and the $\Spfour$ gauge model with two dynamical Dirac flavours. We also
present preliminary results for the meson spectrum in the 
$\Spfour$ gauge theory with two fundamental and three antisymmetric Dirac
flavours. The main motivation of our programme is to test whether this latter model is
viable  as a realisation of Higgs compositeness via the pseudo Nambu
Goldstone mechanism and at the same time can provide partial top compositeness. In this
respect, we report and briefly discuss preliminary results for the
mass of the composite baryon made with two fundamental and one
antisymmetric fermion ({\em chimera baryon}), whose physical properties
are highly constrained if partial top compositeness is at work. Our investigation shows
that a fully non-perturbative study of Higgs compositeness and partial
top compositeness in $\Spfour$ is within reach with our current lattice methodology. 
}
\maketitle
\section{Introduction}
\label{intro}
$\Spfour$ gauge theory with two fundamental and three antisymmetric
Dirac fermion flavours has been suggested as a possible
template~\cite{Barnard:2013zea,Ferretti:2013kya,Cacciapaglia:2019bqz}
for beyond the standard model strong interactions
that can give rise to a composite Higgs boson through the 
breaking of the global fundamental flavour symmetry~\cite{Kaplan:1983fs} and at the same
time to a partial composite top quark state that explains why the
observed mass of the latter particle is at the electroweak
scale~\cite{Kaplan:1991dc}. Partial top compositeness results from the mixing of the
standard model top quark with a chimera baryon, i.e. a baryonic state
formed with two fermions in the fundamental and one fermion in the
antisymmetric representation. A necessary ingredient for partial top
compositeness is the generation of a large anomalous dimension for the
chimera baryon. Lattice gauge theory can be used as a framework to
study non-perturbatively these phenomena, in order to assess their
viability beyond semi-quantitative arguments.

In this work, we shall discuss our previous and current lattice investigations of the composite
Higgs mechanism and of partial top compositeness using $\Spfour$ gauge theory as a
concrete realisation (for studies involving $\SUfour$, see,
e.g.,~\cite{Ayyar:2017qdf,Ayyar:2018zuk,Cossu:2019hse}), starting from
special cases of the target action and approaching the latter in a
{\em crescendo} of complexity. The work is organised as follows. In
Sect.~\ref{sect:2} we introduce the lattice model. Numerical results
are presented in Sect.~\ref{sect:3}. Finally, we draw our conclusions
and we give a brief outlook of forthcoming numerical investigations in
Sect.~\ref{sect:4}. 

\section{Lattice formulation and observables}
\label{sect:2}
The lattice action  we have used in our calculations in $\SptwoN$ gauge
theories with fermions in multiple representations $R$ can be written as
\begin{eqnarray}
S = \beta \sum_x \sum_{\mu<\nu} \left( 1-
  \frac{1}{N_c}\Re \mathrm{Tr} \mathcal{P}_{\mu\nu} \right)  \ 
+ \ \sum_{j = 1}^{N_f^R} \sum_{x,y} \overline{\Psi}_j^R (x) D^{R}
  (x,y) \Psi_j^R(y) \ . 
\label{eq:full_action}
\end{eqnarray}
The first term is the Wilson plaquette action, with $\beta = 2 N_c/g_0^2$, where
$N_c = 2 N$ is the number of colours and $g_0$ is the gauge coupling,
$\mathcal{P}_{\mu\nu}(x)$ is the path exponential of the link variables
$U_{\mu}(x) \in \SptwoN$ around the elementary plaquette stemming from
point $x$ in directions $\mu,\nu$ and $\Re \mathrm{Tr}$ indicates the
real part of the trace. The second term is the fermionic part of the
action. This
term includes a sum over the representations $R$ and over the
flavours $N_f^R$ at fixed representation. For the Dirac operator
$D(x,y)$ we use the Wilson discretisation, given by
\begin{eqnarray}
D^{R}(x,y) = (4+ am_0^{\rm R}) \delta_{x,y}-\frac{1}{2}\sum_\mu
\left\{\frac{}{}(1-\gamma_\mu)U^{R}_\mu(x) \delta_{x+\hat{\mu},y} +
(1+\gamma_\mu)U^{R \dag}_\mu(y) \delta_{x - \hat{\mu},y}\frac{}{}\right\}
\ ,
\label{Eq:DiracF}
\end{eqnarray}
where $U^R_{\mu}(X)$ is the gauge link variable in the representation
$R$, $a$ is the lattice spacing and $m_0^R$ is the bare mass of the
$N_f^R$ degenerate flavours in representation $R$.

The path integral of the theory can be expressed as
\begin{eqnarray}
  Z = \int \left( {\cal D} U \right) \left( {\cal D} \Psi^R \right)
  \left( {\cal D} \overline{\Psi}^R \right) e^{- S} \ , 
\end{eqnarray}
where the integral is performed over all fields and the action, $S$, is given by Eq.~(\ref{eq:full_action}). In this work, we consider two flavours of fermions in the fundamental representation $R \equiv
F$ and three flavours of fermions in the antisymmetric representation
$R \equiv A$.  

Given an operator $O(x_1, \dots, x_n)$, which is a
function of fields at points $x_1, \dots, x_n$, its vacuum expectation
value is computed as
\beq
\langle O(x_1, \dots, x_n)  \rangle = \frac{1}{Z} \int \left( {\cal D} U \right) \left( {\cal D} \Psi^R \right)
  \left( {\cal D} \overline{\Psi}^R \right) O(x_1, \dots, x_n)  e^{-
    S} \ , 
\eeq
where the path integral is calculated using Monte Carlo importance
sampling. 

\begin{figure}[t]
  \begin{center}
    \includegraphics[scale=0.6]{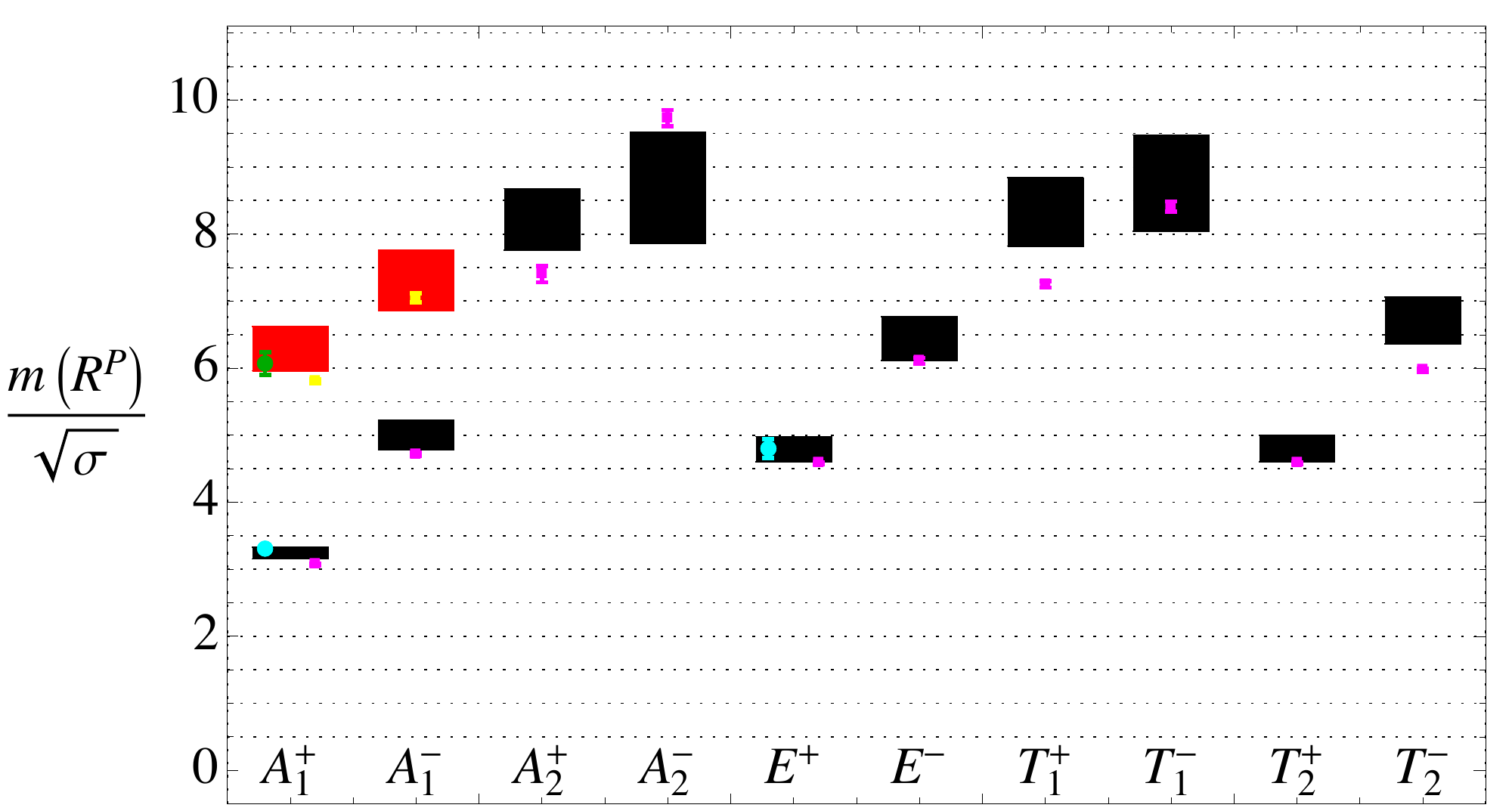}
  \end{center}
\caption{The $\SptwoN$ glueball spectrum in the large-$N$ limit (boxes,
  with ground states in each channel depicted in black and excitations
  in red) compared with $\SUN$ data extrapolated to the same
  limit (circles with error bars). See text for further details. \label{fig:lnspectrum}} 
\end{figure}
\begin{figure}[th]
  \begin{center}
    \includegraphics[scale=0.3]{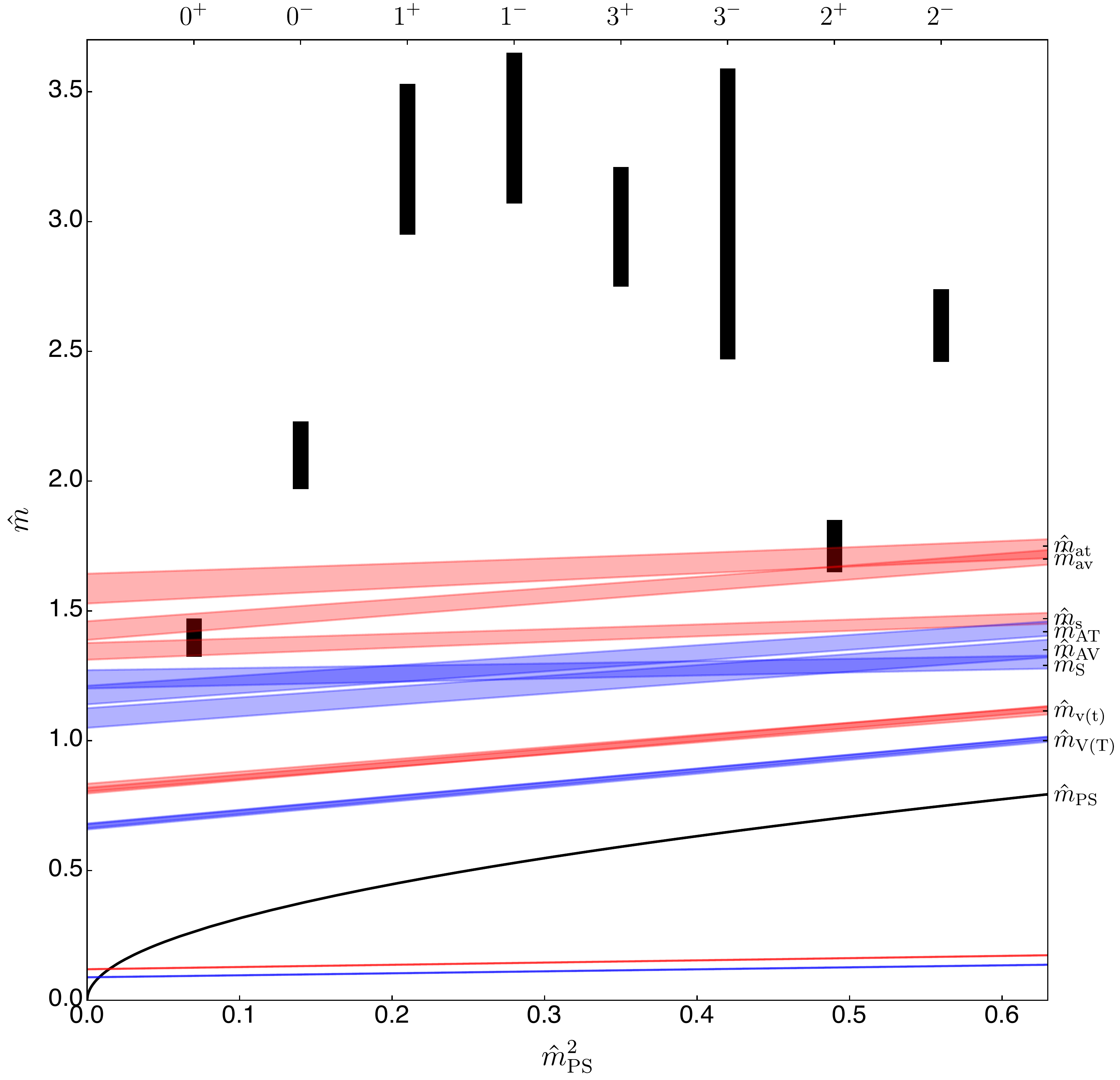}
  \end{center}
  \caption{Quenched spectrum for $\Spfour$ gauge theory with fermions in
    the fundamental and in the antisymmetric
    representation extrapolated to the continuum limit as a function of the pseudoscalar mass
    squared (in black). The bands account for the statistical
    errors. The fundamental spectrum is drown in blue, while the
    antisymmetric spectrum is represented in red. Observables are labelled
    in the right vertical axis. Glueball masses are also reported (boxes), with quantum 
    numbers labelled on the top horizontal axes. Note that, while
    there is a dependency of mesonic quantities from the pion mass,
    since the theory is quenched, the glueball spectrum is constant
    as a function of the constituent fermion mass. \label{fig:quenchedspectrum}}
\end{figure}

We shall consider the following observables:
\begin{itemize}
\item Glueball correlators, which take the form
\beq
\label{eq:gluecorr}
C_{ij}(x_1, x_2) =   \langle O_i(x_1) O^{\dag}_j(x_2) \rangle - \langle O_i(x_1) \rangle \langle O^{\dag}_j(x_2) \rangle \ ,
\eeq
where $O_i(x_1)$ and $O_j(x_2)$ are combinations of Wilson loops
transforming in an irreducible representation of the octahedral group,
at zero momentum and with defined parity;
\item Meson correlators, which are expressed as
\beq
\label{eq:mesoncorr}
C_{\Gamma_1 \Gamma_2} ^R(x_1, x_2) = \left\langle \overline{\Psi}^R (x_1) \Gamma_1
\Psi^R(x_1) \left( \overline{\Psi}^R (x_2) \Gamma_2
\Psi^R(x_2) \right)^{\dag} \right\rangle \ , 
\eeq
with $\Gamma_1$ and $\Gamma_2$ combinations of Euclidean Dirac
matrices, whose specific form determines the $J^{PC}$ quantum numbers of the
states that saturate the propagator;
\item Chimera baryon correlator, written as
  \beq
  C_T^{FFA} (x_1, x_2) =   \left\langle \left( T \Psi ^F (x_1)  \Psi ^F
      (x_1)  \Psi ^A (x_1)  \right)  \left( T \Psi^F (x_2)  \Psi^F
      (x_2)  \Psi ^A (x_2)  \right) ^\dag \right \rangle \ , 
  \eeq
where $T$ is a tensor contracting flavour and colour indices, which, for the sake of conciseness,
will be left implicit. More details on
this construction will be provided in a forthcoming publication.  
\end{itemize} 

\section{Numerical Results}
\label{sect:3}
Before tackling the full theory, relevant operators have been studied
in notable limits of the latter. Since the properties of the model in
these limits are better known, the simplified calculations have
enabled us to develop the methodology and at the same time to provide
a set of safe reference measurements against which to check the
complete calculations. Here, we report a selection of our numerical
results. For additional measurements and their discussion, we refer to the
quoted original papers. 

The first model we have considered is the pure gauge system (first
investigated in~\cite{Holland:2003kg}). In this case, relevant observables are glueball masses and the string tension. These
are extracted from correlators of the type given in Eq.~(\ref{eq:gluecorr}),
looking at the asymptotic behavior
\beq
C_{ij}(x_1, x_2) \mathop{\simeq}_{|x_1 - x_2 | \to \infty} C e^{- M
  |x_1 - x_2|} \ ,
\eeq
where $M$ is the mass of the lowest-lying state carrying the quantum
numbers of the source operators. In practice, the calculation of single
correlators is very noisy. The bad signal-to-noise ratio can be
overcome with a thorough variational calculation. The methodology we
have used is explained in~\cite{Bennett:2020hqd}. The string tension
can be extracted similarly, considering correlators of Polyakov
loops. 

We have computed in~\cite{Bennett:2020hqd} the lowest-lying glueball spectrum in units of the
string tension in the continuum limit for $\SptwoN$ models with $N = 2, 3, 4$
and we have extrapolated the latter to $N \to \infty$. For the
large-$N$ extrapolation, we have found that the leading expected $1/N$
correction provides a good description of the data. Our results are displayed by the boxes of
Fig.~\ref{fig:lnspectrum}. The black boxes indicate the masses of the
groundstate in each channel, with the red boxes representing masses of
the excited states (for the channels for which the latter turned out
to be measurable on our samples). The quantum numbers refer to the irreducible
representations of the octahedral group. The continuum $J^{P}$ quantum
numbers\footnote{We remind the reader that, since $\SptwoN$ groups are
  pseudoreal, the charge conjugation quantum number of glueballs in $\SptwoN$
  is always $+1$.} can be reconstructed following the method
described, e.g., in~\cite{lucini:2010nv}.

It is worth remarking that, since the large-$N$ limit of $\SUN$ and
$\SptwoN$ coincide, for the majority of the glueball states, our work is the first
extrapolation of the glueball spectrum to $N = \infty$. Indeed, the only states
whose $\SUN$ large-$N$ mass had been determined before our calculations are
the groundstate and the first excitation in the spin zero channel (in
our notation, the $A_1$ channel) and the spin two (which appears in
both the $E$ and the $T_2$ channel)~\cite{Lucini:2001ej,Lucini:2004my}
(see also~\cite{Lucini:2012gg}). We report in
Fig.~\ref{fig:lnspectrum} the more constraining measurements
of~\cite{Lucini:2004my} (in cyan for the groundstates and in green for
the excitation in the $A_1^+$ channel), which are in full agreement
with ours. After our work was published, a comprehensive study of the
large-$N$ $\SUN$ glueball spectrum was
published~\cite{Athenodorou:2021qvs}. This enables us to perform a broader 
comparison of the two extrapolations. The results of~\cite{Athenodorou:2021qvs}
are also reported in Fig.~\ref{fig:lnspectrum} (in magenta for the
groundstates, and in yellow for the excitations). While we notice that
the authors of~\cite{Athenodorou:2021qvs} quote significantly smaller errors, the
agreement with our measurements is within the error bars (as
measured on our own data) in most of the cases, with the largest
discrepancies being at most around two standard deviations. This
comparison provides evidence of the universality of the
large-$N$ glueball spectrum across the $\SUN$ and the $\SptwoN$
series. 

\begin{figure}[t]
  \begin{center}
    \includegraphics[scale=0.6]{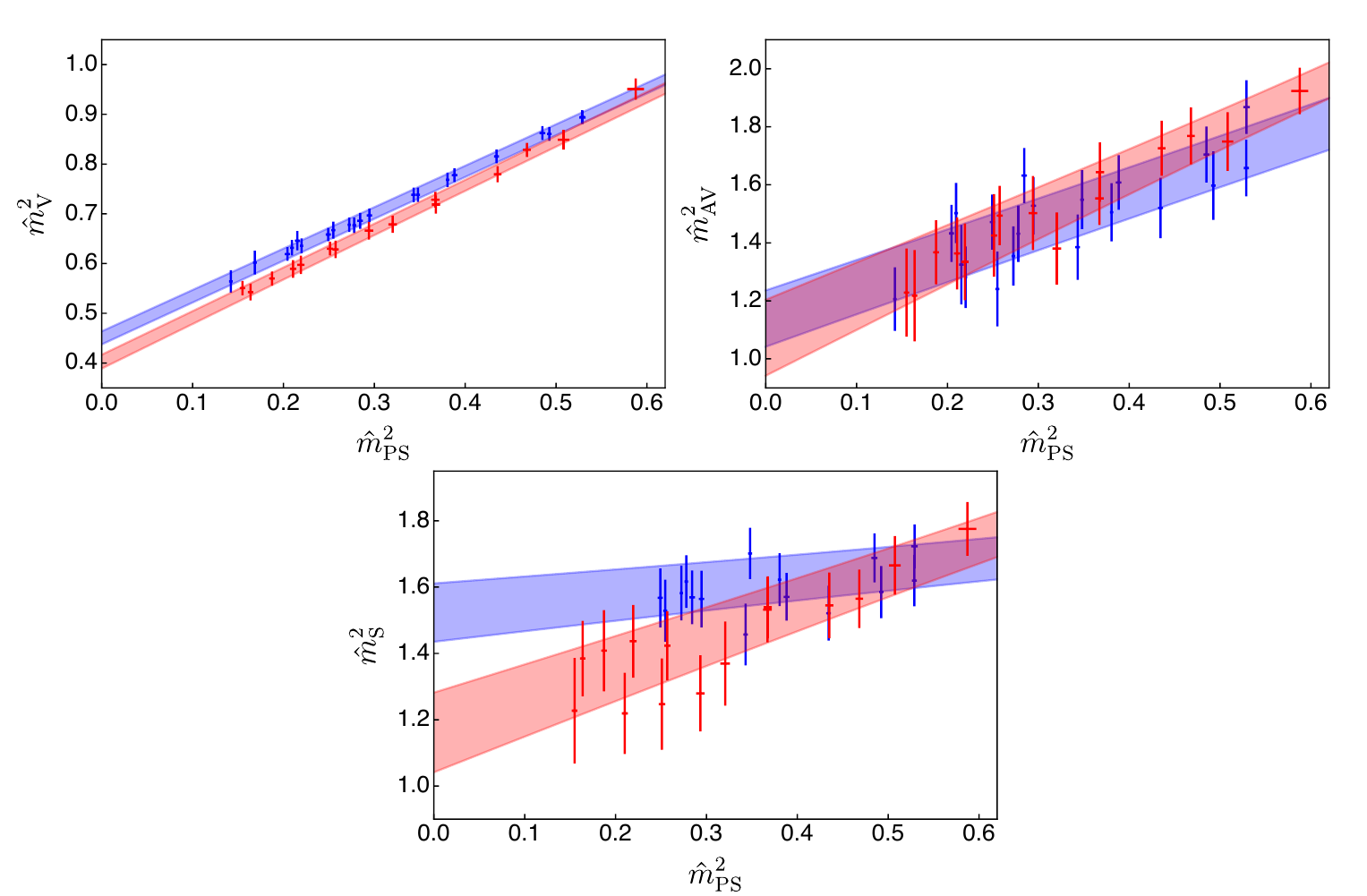}
  \end{center}
 \caption{Squared masses of the vector (top left), the axial vector
   (top right) and the isotriplet scalar meson (bottom) as a function of the squared mass of the
   pseudoscalar for the fundamental representation in the continuum limit. We plot in red dynamical data and in blue quenched data, with the corresponding bands representing
   the chiral extrapolation. \label{fig:comparison_mmasses}}
\end{figure}

We now discuss the quenched spectrum of the $\Spfour$ gauge theory, comparing masses
of glueballs and mesons, the latter studied as the mass of the
constituent fermions is decreased. We consider a mesonic
correlators of the form~(\ref{eq:mesoncorr}). Its asymptotic behaviour is given
by
\beq
C_{\Gamma_1 \Gamma_2} ^F(x_1, x_2) \simeq A \left( e^{- M t} + e^{- M (T -
    t) } \right)  \ ,
\eeq
where the second term (due to waves travelling across the antiperiodic
boundary of our lattice) has now been inserted explicitly, as, unlike
in the glueball case, neglecting it would give a measurable systematic
error.  In the previous formula, $T$ is the temporal extension of the
lattice. Points $x_1$ and $x_2$  have spacial coordinates summed over
$|\vec{x}_1-\vec{x}_2$| at fixed $\vec{x}_2$ (i.e., the correlator has
zero net momentum). The temporal coordinates are respectively $\tau$ and $t + \tau$ 
at fixed $\tau < T$. The decay constant of the
considered states can be extracted from $A$, either directly or (in
the case of the pseudoscalar meson) considering two appropriate
combinations of correlators. We refer to the quoted original
works for the details, including renormalisation of decay
constants and how the choice of the $\Gamma$ matrices identifies the
$J^{PC}$ quantum numbers.

The continuum quenched meson spectrum (in both the fundamental and antisymmetric
representation), originally discussed in~\cite{Bennett:2019cxd}, is
reported in Fig.~\ref{fig:quenchedspectrum}, together with a selection
of glueball states taken from~\cite{Bennett:2020qtj}. The scale is set
using the gradient flow derived quantity $w_0$, as described
in~\cite{Bennett:2017kga}. The 
meson spectrum in both representations is plotted as a function of the pseudoscalar
mass in that representation. Glueball states are also reported for
the same model. The data show some clear indications: (1) in both
representations, the general features of the quenched spectrum are
qualitatively similar to those of QCD; (2) while the
meson spectrum has the same behaviour in both representations, at fixed pseudoscalar mass
antisymmetric states are around 10-20\% heavier than the corresponding
fundamental states, with the percent difference increasing towards the light
fermion regime; (3) glueballs are generally significantly heavier than
mesons, but the lightest glueball states have masses that are comparable with masses of states
in the heaviest mesonic channels. 

We have then studied the theory with fermionic matter, comparing the
two-flavour dynamical model~\cite{Bennett:2019jzz} to the quenched
case~\cite{Bennett:2017kga} for fundamental representation fermions in the
continuum limit. The masses of the vector, axial vector and isotriplet
scalar meson are plotted in Fig.~\ref{fig:comparison_mmasses}. With
the exception of the latter, typically unquenching effects are very
small, and hardly visible in the range of the obtained pion masses. 

\begin{figure}[t]
  \begin{center}
    \includegraphics[scale=0.3]{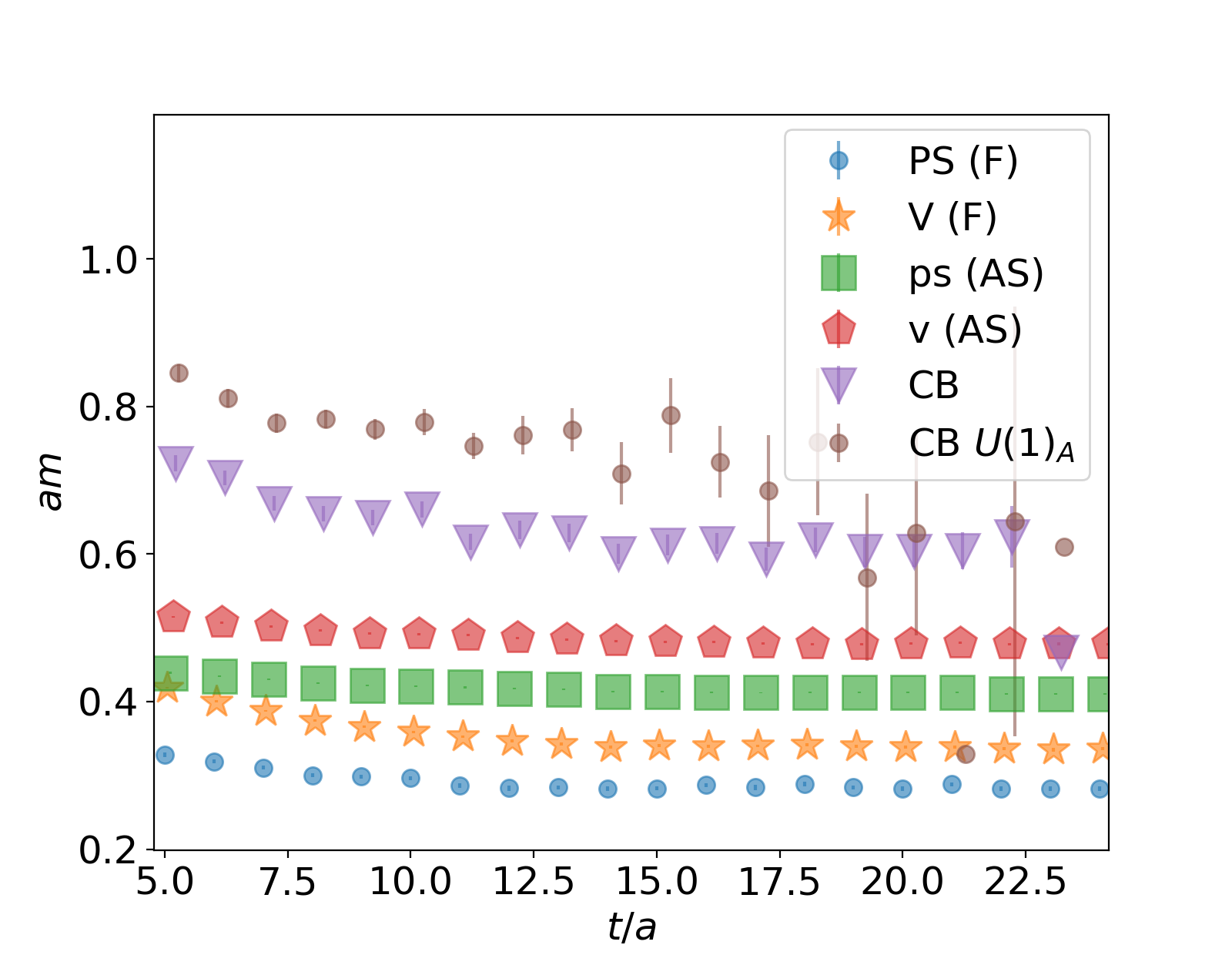}
      \includegraphics[scale=0.3]{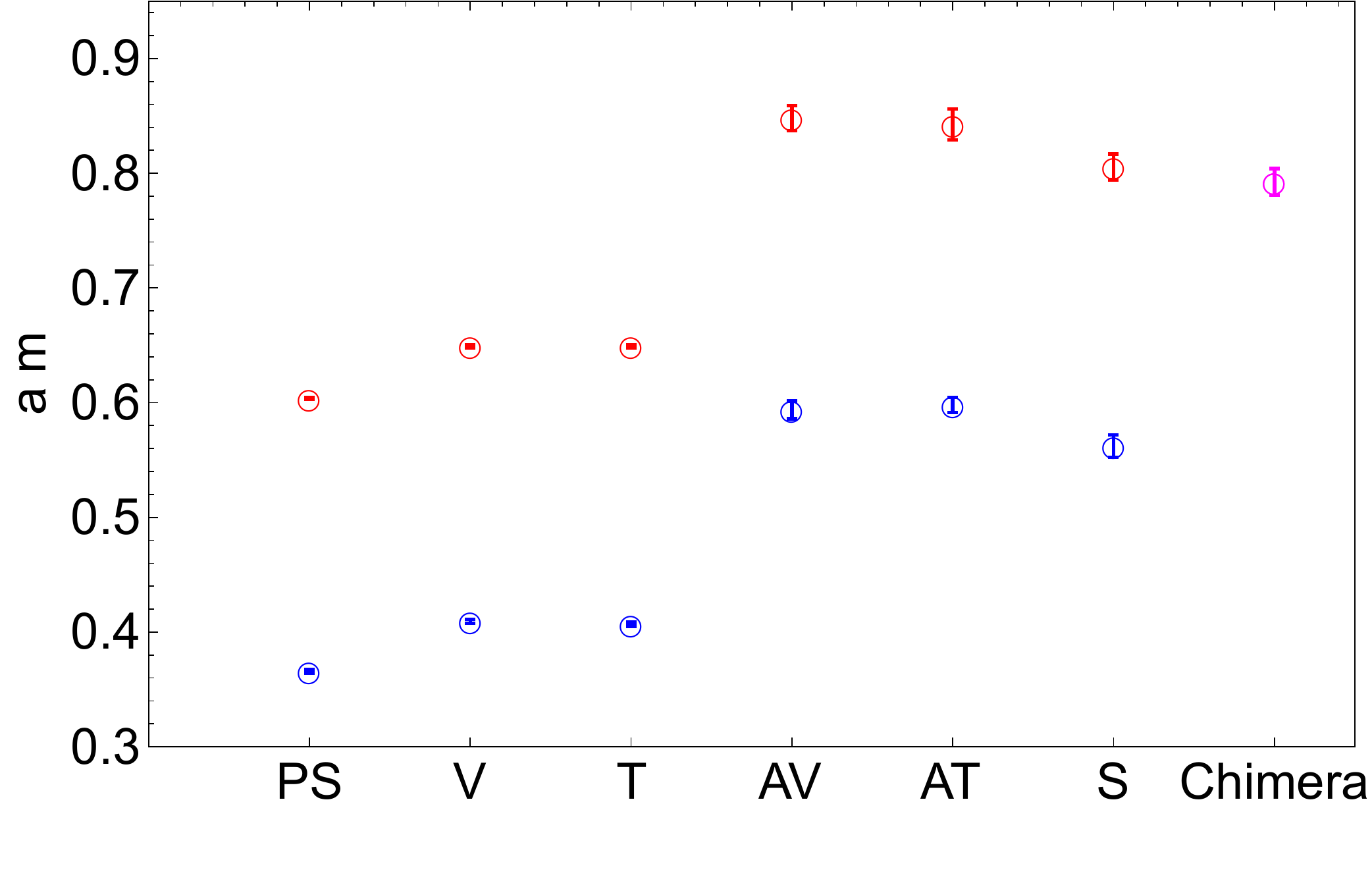}
    \end{center}
    \caption{Left: The lowest-lying meson spectrum in $\Spfour$ with three dynamical antisymmetric fermions. Taking quenched fundamental sources in this model, we also show the pseudoscalar and the vector meson masses in the fundamental representation, the chimera baryon (CB) and its $U(1)_A$ partner. The lattice parameters are discussed in the text. Right: The meson spectrum for both representations and the chimera baryon in the $\Spfour$ gauge theory with two fundamental dynamical flavours and three antisymmetric dynamical flavours, at the lattice parameters discussed in the text. \label{fig:mixed_baryon}}
  \end{figure}

In the subsequent stage of our investigation programme of $\Spfour$ gauge models, we
have studied the system in a partially quenched setup, considering three
dynamical antisymmetric fermions and two static fundamental
flavours. This gives us an opportunity to perform a first study of the
chimera baryon in a model that is less complicated to simulate than
the target mixed-representation system. Figure~\ref{fig:mixed_baryon} (left) displays our preliminary
results for the two lowest-lying
mesonic states (the pseudoscalar and the scalar) in both
representations, together with the chimera baryon and its parity
partner. Results have been obtained on a $48 \times 24^3$ lattice
at gauge coupling $\beta = 6.65$, antisymmetric fermion bare mass
$a m_0^A = -1.07$ and quenched fundamental mass $a m_0^F =
-0.734$. Statistical noise has been reduced using
Wuppertal smearing~\cite{Gusken:1989qx}. The data show that in our theory and
for the used lattice parameters, chimera baryons are heavier
than the antisymmetric vector meson. Beyond the (not unimportant) numerical details, likely to 
be dependent on our particular setup, the main conclusion of our
investigation is that the technology we have developed enables us to
determine masses of chimera baryons. Note in particular the visible
splitting between the chimera baryon and its $U(1)_A$ partner. 

Finally, we are now performing calculations in $\Spfour$ gauge theory
with two dynamical fundamental fermions and three dynamical
antisymmetric fermions. A preliminary plot the meson spectrum
and of the chimera baryon is reported in Fig.~\ref{fig:mixed_baryon},
right. The lattice parameters are $\beta = 6.5$, $a m_0^A = -1.01$ and
$a m_0^F  = -0.71$, and the lattice size is $48 \times 24^3$. The
results show that for our choice of parameters the chimera 
baryon has a mass comparable to that of the axial vector and of the
isotriplet scalar mesons in the antisymmetric representation. 
\section{Conclusions and outlook}
\label{sect:4}
Following the programme we started in~\cite{Bennett:2017kga}, we have
performed a set of extensive simulations aimed to assess the viability
of $\Spfour$ gauge theory with two fundamental and three antisymmetric
fermions as a realisation of the composite Higgs and of the partial
top compositeness mechanism. So far we have developed the needed
technology by studying the glueball spectrum~\cite{Bennett:2020qtj}, the quenched
meson spectrum~\cite{Bennett:2019cxd} and the dynamical theories respectively with
two fundamental~\cite{Bennett:2019jzz} and three antisymmetric fermion
flavours. A selection of the results is reported in this work.  
A more systematic calculation in the target phenomenological
model, which includes dynamical fermions in both representations, is  now within
reach. We have presented some preliminary results for the
latter model. A more detailed account of its features will be reported
in a forthcoming publication. Another interesting direction, pursued by
the authors of~\cite{Maas:2021gbf}, is to
explore an $\Spfour$ gauge models with two non-degenerate fundamental
fermions in the context of SIMP dark matter.\\
~\\
\begin{acknowledgement}
\noindent
{\bf Acknowledgements -} The work of E. B. has bee n funded by the Supercomputing Wales project,
which is part-funded by the European Regional Development Fund (ERDF)
via Welsh Government. J. H. is supported by the STFC Consolidated
Grant No. ST/P00055X/1, by the College of Science,
Swansea University, and by the Grant No. STFC-DTG
ST/R505158/1. The work of D. K. H. was supported by
Basic Science Research Program through the National
Research Foundation of Korea (NRF) funded by the
Ministry of Education (NRF-2017R1D1A1B06033701).
The work of J. W. L. is supported in part by the National
Research Foundation of Korea funded by the Ministry of
Science and ICT (NRF-2018R1C1B3001379) and in part
by Korea Research Fellowship program funded by the
Ministry of Science, ICT and Future Planning through
the National Research Foundation of Korea
(2016H1D3A1909283). The work of C. J. D. L. is supported
by the Taiwanese MoST Grant No. 105-2628-M-009-003-
MY4. The work of B. L. and M. P. has been supported in part
by the STFC Consolidated Grant No. ST/T000813/1. B. L. and M. P. received funding from 
the European Research Council (ERC) under the European
Union’s Horizon 2020 research and innovation program
under Grant Agreement No. 813942. The work of B. L. is
further supported in part by the Royal Society Wolfson
Research Merit Award No. WM170010 and by the
Leverhulme Trust Research Fellowship No. RF-2020-4619. The work of
D. V. is supported in part by the INFN HPCHTC project and in part by
the Simons Foundation under the 
program “Targeted Grants to Institutes” awarded to the
Hamilton Mathematics Institute. Numerical
simulations have been performed on the Swansea SUNBIRD
cluster (part of the Supercomputing Wales project) and AccelerateAI A100 GPU system, on the local HPC clusters 
in Pusan National University (PNU) and in National Chiao-Tung University (NCTU), and on the Cambridge Service for Data Driven
Discovery (CSD3). The Supercomputing Wales project and AccelerateAI are part funded
by the European Regional Development Fund (ERDF) via
Welsh Government. CSD3 is operated in part by the
University of Cambridge Research Computing on behalf
of the STFC DiRAC HPC Facility (www.dirac.ac.uk). The
DiRAC component of CSD3 was funded by BEIS capital
funding via STFC capital Grants No. ST/P002307/1 and
No. ST/R002452/1 and STFC operations Grant No. ST/R00689X/1. DiRAC is
part of the National e-Infrastructure. 
\end{acknowledgement}

\bibliography{sp4}

\begin{thebibliography}{21}

\bibitem{Barnard:2013zea}
J.~Barnard, T.~Gherghetta, T.S. Ray, JHEP \textbf{02}, 002 (2014),
  \texttt{1311.6562}

\bibitem{Ferretti:2013kya}
G.~Ferretti, D.~Karateev, JHEP \textbf{03}, 077 (2014), \texttt{1312.5330}

\bibitem{Cacciapaglia:2019bqz}
G.~Cacciapaglia, G.~Ferretti, T.~Flacke, H.~Ser\^odio, Front. in Phys.
  \textbf{7}, 22 (2019), \texttt{1902.06890}

\bibitem{Kaplan:1983fs}
D.B. Kaplan, H.~Georgi, Phys. Lett. B \textbf{136}, 183 (1984)

\bibitem{Kaplan:1991dc}
D.B. Kaplan, Nucl. Phys. B \textbf{365}, 259 (1991)

\bibitem{Ayyar:2017qdf}
V.~Ayyar, T.~DeGrand, M.~Golterman, D.C. Hackett, W.I. Jay, E.T. Neil,
  Y.~Shamir, B.~Svetitsky, Phys. Rev. D \textbf{97}, 074505 (2018),
  \texttt{1710.00806}

\bibitem{Ayyar:2018zuk}
V.~Ayyar, T.~Degrand, D.C. Hackett, W.I. Jay, E.T. Neil, Y.~Shamir,
  B.~Svetitsky, Phys. Rev. D \textbf{97}, 114505 (2018), \texttt{1801.05809}

\bibitem{Cossu:2019hse}
G.~Cossu, L.~Del~Debbio, M.~Panero, D.~Preti, Eur. Phys. J. C \textbf{79}, 638
  (2019), \texttt{1904.08885}

\bibitem{Holland:2003kg}
K.~Holland, M.~Pepe, U.J. Wiese, Nucl. Phys. B \textbf{694}, 35 (2004),
  \texttt{hep-lat/0312022}

\bibitem{Bennett:2020hqd}
E.~Bennett, J.~Holligan, D.K. Hong, J.W. Lee, C.J.D. Lin, B.~Lucini, M.~Piai,
  D.~Vadacchino, Phys. Rev. D \textbf{102}, 011501 (2020), \texttt{2004.11063}

\bibitem{lucini:2010nv}
B.~Lucini, A.~Rago, E.~Rinaldi, JHEP \textbf{08}, 119 (2010),
  \texttt{1007.3879}

\bibitem{Lucini:2001ej}
B.~Lucini, M.~Teper, JHEP \textbf{06}, 050 (2001), \texttt{hep-lat/0103027}

\bibitem{Lucini:2004my}
B.~Lucini, M.~Teper, U.~Wenger, JHEP \textbf{06}, 012 (2004),
  \texttt{hep-lat/0404008}

\bibitem{Lucini:2012gg}
B.~Lucini, M.~Panero, Phys. Rept. \textbf{526}, 93 (2013), \texttt{1210.4997}

\bibitem{Athenodorou:2021qvs}
A.~Athenodorou, M.~Teper (2021), \texttt{2106.00364}

\bibitem{Bennett:2019cxd}
E.~Bennett, D.K. Hong, J.W. Lee, C.J.D. Lin, B.~Lucini, M.~Mesiti, M.~Piai,
  J.~Rantaharju, D.~Vadacchino, Phys. Rev. D \textbf{101}, 074516 (2020),
  \texttt{1912.06505}

\bibitem{Bennett:2020qtj}
E.~Bennett, J.~Holligan, D.K. Hong, J.W. Lee, C.J.D. Lin, B.~Lucini, M.~Piai,
  D.~Vadacchino, Phys. Rev. D \textbf{103}, 054509 (2021), \texttt{2010.15781}

\bibitem{Bennett:2017kga}
E.~Bennett, D.K. Hong, J.W. Lee, C.J.D. Lin, B.~Lucini, M.~Piai, D.~Vadacchino,
  JHEP \textbf{03}, 185 (2018), \texttt{1712.04220}

\bibitem{Bennett:2019jzz}
E.~Bennett, D.K. Hong, J.W. Lee, C.J.D. Lin, B.~Lucini, M.~Piai, D.~Vadacchino,
  JHEP \textbf{12}, 053 (2019), \texttt{1909.12662}

\bibitem{Gusken:1989qx}
S.~Gusken, Nucl. Phys. B Proc. Suppl. \textbf{17}, 361 (1990)

\bibitem{Maas:2021gbf}
A.~Maas, F.~Zierler, PoS(LATTICE2021) to appear  (2021), \texttt{2109.14377}

\end{thebibliography}

\end{document}